# GIANT ATMOSPHERIC SHOWERS DETECTED BY THE YAKUTSK EXTENSIVE AIR SHOWERS ARRAY

A. V. Glushkov[1], K. G. Lebedev[1], L. T. Ksenofontov[1], A. V. Saburov, O. N. Ivanov[1], A. F. Boyakinov[1], A. A. Ivanov[1], S. P. Knurenko[1], A. D. Krasilnikov[1], S. V. Matarkin[1], V. P. Mokhnachevskaya[1], N. Ya. Muksunov[1], I. S. Petrov[1], I. E. Sleptsov[1]

[1]Yu.G. Shafer Institute of Cosmophysical Research and Aeronomy of the Siberian Branch of the Russian Academy of Sciences, Yakutsk, Russia

*E-mail: glushkov@ikfia.ysn.ru

The two most powerful extensive air showers (EAS) with energies of about $10^{20}$ eV, registered at the Yakutsk EAS array during the entire observation period of 1974-2024, are considered. Both showers hit the array near the center and triggered all surface detectors and underground muon detectors with a threshold energy of $E_\mu = 1.0 \times \cos\theta$ GeV. These events have an abnormally high fraction of muons, which is beyond current model predictions. This may change our understanding of hadron interactions at ultra-high energy, but there is also a possibility that these showers were initiated by some exotic primary particles.

## INTRODUCTION

Extensive air showers (EAS) of extreme energies ($E \approx 10^{20}$ eV) has always been [1–5] and continues to be [6–8] of great interest. These events are very rare and can be registered only by largest world EAS arrays. After the discovery of cosmic microwave background (CMB) it was established that due to interaction of primary protons and nuclei with CMB, the cosmic ray (CR) flux must sharply decrease at $E > 5 \times 10^{19}$ eV – the GZK-effect [9,10]. This imposes certain restrictions on the distance between Earth and the sources of generation of CR with these energies. It was reported that on May 27 2021 the Telescope Array observatory had registered a primary particle with maximum observed energy $E = (2.44 \pm 0.29) \times 10^{20}$ eV [8]. This unique event was called "Amaterasu". There are several EAS events of extreme energy registered at the Yakutsk EAS array.

## EAS EVENTS OF EXTREME ENERGY

Below we examine two giant EAS events whose axes lie near the array center. They provided important empirical information on the lateral distribution of cascade particles in such rare showers. These EASs have triggered all currently operating surface detectors (SD) and underground muon detectors (MD) with threshold energy $E_\mu = 1.0 \times \cos\theta$ GeV. Additional point of interest in these showers is their unusually high muon content. Table 1 lists some of their main characteristics. Column 3 lists arrival zenith angles. In columns 4 and 5 densities of SD and MD response are indicated at axis distance $r = 600$ m determined with the use of lateral distribution functions (LDF):

$$S(r,\theta) = S_{600}(\theta) \times \left(\frac{600}{r}\right)^2 \times \left(\frac{600 + r_0}{r + r_0}\right)^{\beta(\theta)-2} \times \left(\frac{600 + r_1}{r + r_1}\right)^{10}, \quad (1)$$

with $r_0 \approx 70$ m and $r_1 = 10^4$ m;

**Table 1.** Parameters of giant EAS events.

| 1 | 2 | 3 | 4 | 5 | 6 | 7 | 8 | 9 |
|---|---|---|---|---|---|---|---|---|
| Date | log($E$/eV) | $\theta$ (deg.) | log($S_{600}(\theta)$/m$^{-2}$) | log($S_{\mu,600}(\theta)$/m$^{-2}$) | RA | DEC | $L_G$ | $B_G$ |
| | | | | | (deg.) | | | |
| 07/05/1989 | 20.05 | 58.7 | 1.74 ± 0.02 | 1.81 ± 0.04 | 75.0 | 45.6 | 162.0 | 2.6 |
| 02/04/2024 | 19.81 | 56.9 | 1.58 ± 0.02 | 1.48 ± 0.04 | 142.7 | 45.6 | 174.8 | 47.0 |

$$S_\mu(r,\theta) = S_{\mu,600}(\theta) \times \left(\frac{600}{r}\right)^{0.75} \times \left(\frac{600 + r_2}{r + r_2}\right)^{\beta(\theta)-2} \times \left(\frac{600 + r_3}{r + r_3}\right)^{9}, \quad (2)$$

with $r_2 = 280$ m and $r_3 = 2\times10^3$ m [11]. The parameters $S_{600}(\theta)$, $\beta(\theta)$, $S_{\mu,600}(\theta)$, $b_\mu(\theta)$ were determined from the measured densities from $\chi^2$-minimization during the preliminary events processing. In columns 6, 7 and 8, 9 corresponding equatorial galactic coordinates are listed.

*Event from 07/05/1989*

First shower, whose energy at the time of registration was estimated as $E \approx 1.2\times10^{20}$ eV [5], has turned out to be the most powerful event in the entire operation of the Yakutsk array. It was dubbed "Arian". Its energy was later re-evaluated to a higher value, $E = (1.7 - 2.0) \times10^{20}$ eV [12]. The modern estimation is $E \approx 1.1\times10^{20}$ eV [13]. The layout of scintillation SDs and MDs for the period of registration of this shower is shown on Fig. 1. The geometrical area of the array was 17.32 km$^2$. Two MDs had an area 36 m$^2$ each and had been operational since 1976. They were located at ≈ 500 m from the array center. In 1986 three additional MDs were deployed with an area 20 m$^2$ each. Sizes of circles on Fig. 1 are proportional to the logarithm of summary registered SD response. Empty circles indicate observational stations that were put offline for the maintenance.

LDFs for response densities of SD and MD are shown on Fig. 2. Light circles indicate readings from six additional stations in the central area with condensed spacing 250 m. Each station contained one standard 2-m$^2$ scintillation detector. Additionally, a compact cluster of 0.25-m$^2$ scintillation detectors with 50 m spacing was operational in the central area of array. Solid curve corresponds to best fit of approximation (1) to readings of all SDs. It has turned out to be close to

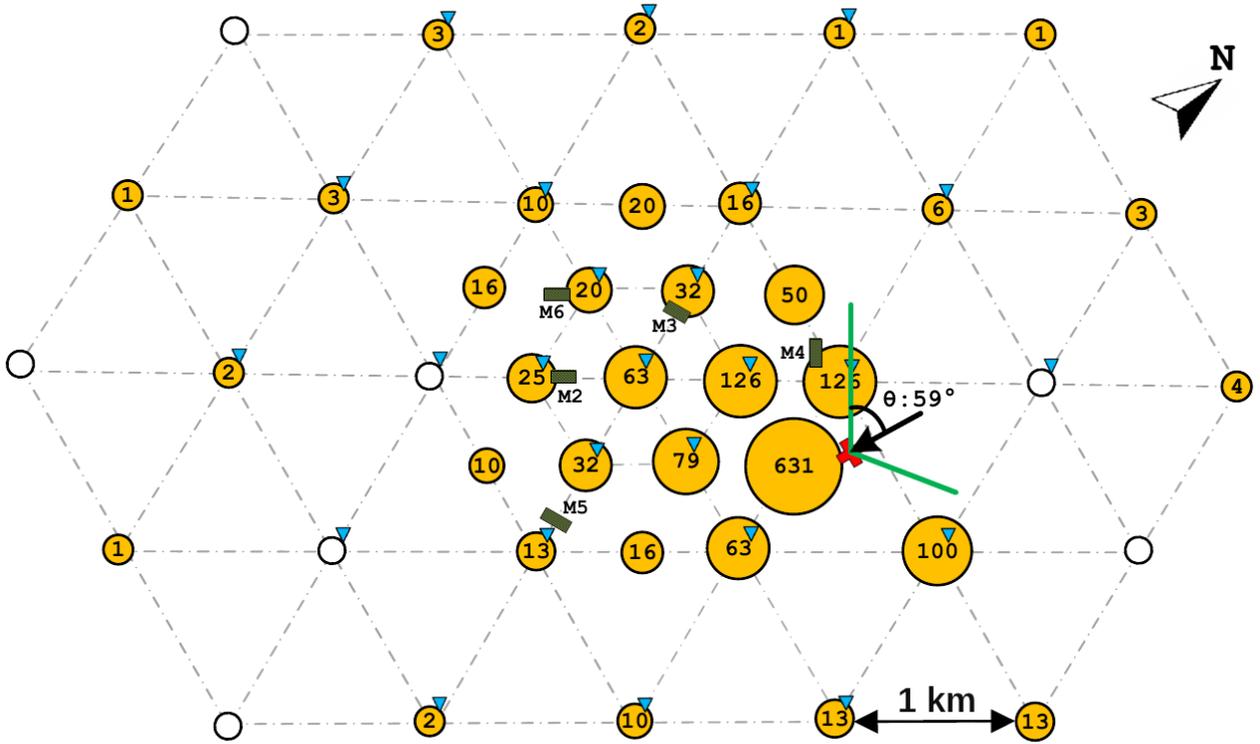

**Figю 1.** Footprint of the Arian shower on the array plane. Numbers reflect summary SD responses. Empty circles represent observational stations put offline, dark squares − MDs with $E_\mu = 1.0 \times \sec\theta$ threshold, triangles − detectors of Cherenkov light from EAS. Axis coordinates and shower arrival direction are denoted with cross and arrow.

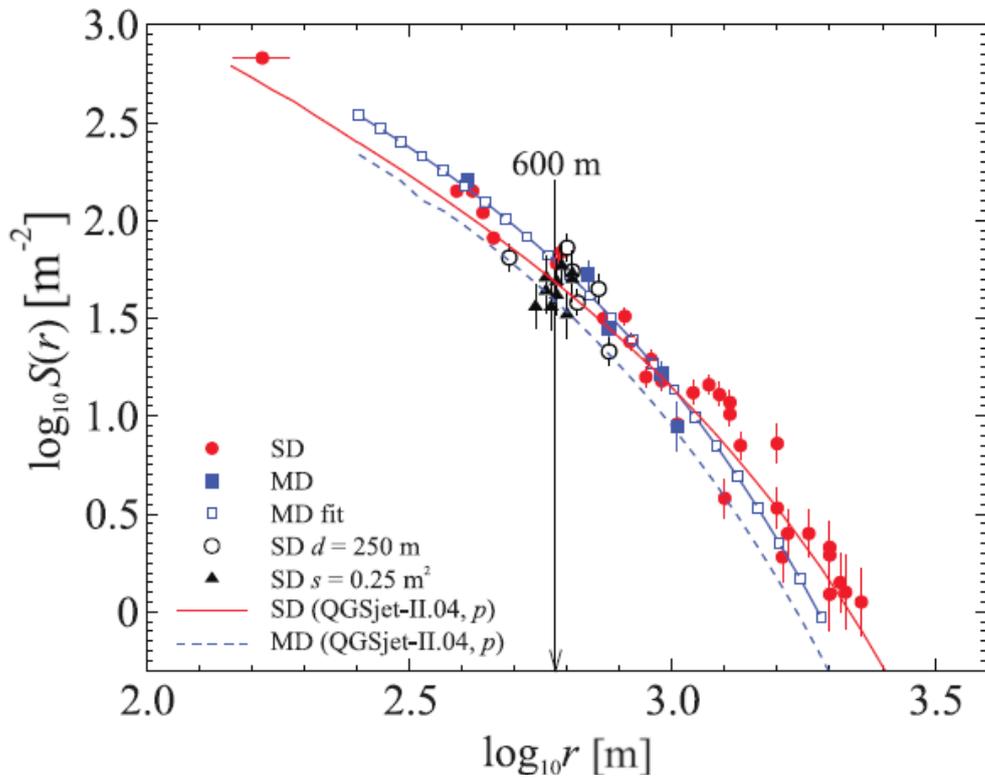

**Fig. 2.** Lateral distributions of particles in the Arian event at observation level. Circles and triangles represent all-particle distribution, dark squares −distribution of muons with energies above $1.0 \cdot \sec\theta$ GeV. Solid and dashed curves − results simulation performed within framework of QGSJet-II.04 model for primary protons with energy $10^{20}$ eV и $\theta = 59°$. The curve with empty squares − approximation (2) of the experimental data.

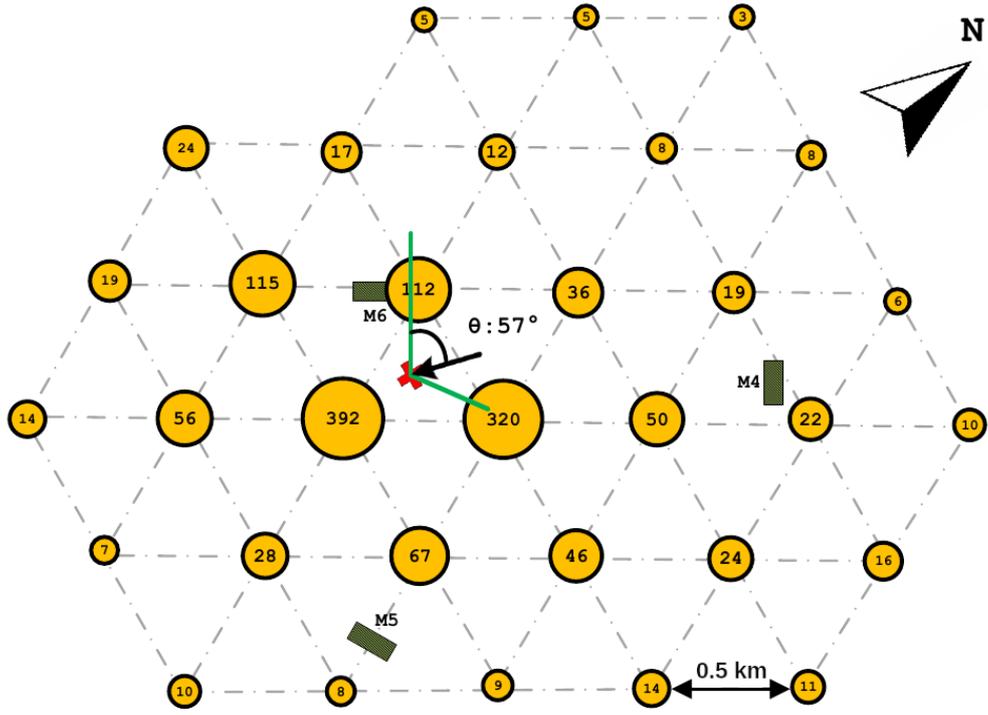

**Fig. 3.** Footprint of the EAS with energy 6.46×10$^{19}$ eV on the array plane. The symbols are the same as on Fig. 1.

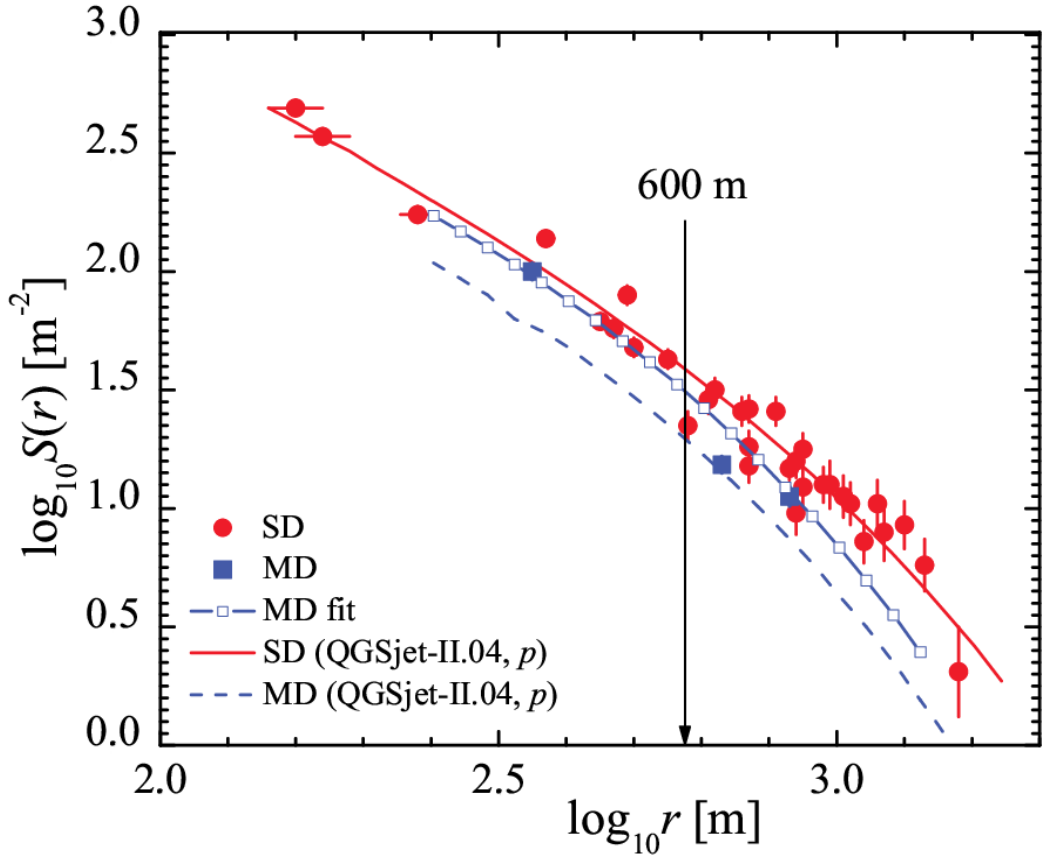

**Fig. 4.** Lateral distributions of EAS particles in the event from 02/04/2024. Solid and dashed lines – simulation results for SD and MD according to QGSJet-II.04 model. The curve with empty squares – approximation (2) of experimental data.

**Table 2.** Response densities from muons (evts/m2) in separate scintillation counters (MD, see Fig. 3) at axis distance r(m); <$S_\mu(r,\theta)$> − mean density.

| MD | 1 | 2 | 3 | 4 | 5 | 6 | 7 | 8 | 9 | 10 | r | <$S_\mu(r,\theta)$> |
|---|---|---|---|---|---|---|---|---|---|---|---|---|
| M4 | 16.7 | 21.0 | 20.7 | 16.0 | 7.9 | 16.4 | 7.0 | 9.4 | 9.3 | 28.6 | 682 | 15.3 |
| M5 | 11.3 | 9.4 | 8.0 | - | 7.1 | 11.3 | 15.3 | 17.1 | 10.6 | 15.0 | 850 | 11.7 |
| M6 | 93.3 | 97.4 | 84.1 | 62.6 | 130.1 | 111.2 | 92.9 | 125.4 | 108.8 | 85.2 | 353 | 99.2 |

the LDF obtained within the framework of the QGSJet-II.04 model [14] for primary protons with energy $10^{20}$ eV and $\theta = 59°$ in both absolute value and shape. The curve with empty squares represents the lateral distribution (2) of experimentally measured responses of MDs with threshold energy $E_\mu \geq 1.92$ GeV. According to the $\chi^2$-test, this approximation agrees with experimental data with the value $\log_{10}[S^{exp}_{\mu,600}(E, 59°)] = 1.81 \pm 0.04$ m$^{-2}$ (column 5 in Table 1). Corresponding muon densities were calculated with the use of QGSjet-II.04 model. Their values are $\log_{10}[S_{\mu,600}(E, 59°, p)] = 1.51 \pm 0.02$ for primary protons (Fig. 2) and $\log_{10}[S_{\mu,600}(E, 59°, Fe)] = 1.66 \pm 0.02$ for iron nuclei.

*Event from 02/04/2024*

During the renovation that took place in 1990-1992, the farthermost SDs (see Fig. 1) were demount and moved into a central circle of a 2 km radius. In 2010 the area of the array was reduced to 8.23 km$^2$. During the 2019-2021 period a significant technical modernization was conducted, and as a result the array was reduced to its minimal area of 5.62 km$^2$. The geometry of the Yakutsk array at the moment of registration of second giant shower is shown on Fig. 3. It is evident from Table 1 that the second giant shower has triggered nearly all SDs and MDs. This reflects the high quality of the revised and modernized array electronics. The readings of all muon counters that demonstrate coordinated operation of three MDs are listed in Table 2. By that time two other 36-m$^2$ MDs have been decommissioned due to failures of their underground housings. The densities of SD and MD responses recorded in this shower are shown on Fig. 4. Solid curve represents best fit of the approximation (1) to SD readings. It is also very close in absolute value and shape to the LDF obtained with the use of QGSJet-II.04 model for primary protons with energy 6.46×10$^{19}$ eV and $\theta$ = 57°. The curve with empty squares represents the LDF (2) of measured responses of MD with $E_\mu \geq 1.83$ GeV threshold. It fits the experimental data with the value $\log_{10}[S^{exp}_{\mu,600}(E, 57°)] = 1.48 \pm 0.02$ m$^2$ presented in the column 5 of Table 1. Corresponding values obtained for this shower with the framework of QGSJet-II.04 model are $\log_{10}[S_{\mu,600}(E, 57°, p)] = 1.30 \pm 0.02$ for primary protons (Fig. 4) and $\log_{10}[S_{\mu,600}(E, 57°, Fe)] = 1.37 \pm 0.02$ for iron nuclei.

## DISCUSSION

The data presented on Fig. 2 and Fig. 4 were obtained during different periods of the Yakutsk array's operation, with significantly different technical abilities of the instrument. It is evident from this figures that the muon fractions in showers discussed above are very close. For example, their values at axis distance $r = 600$ m in both showers are $\frac{S^{exp}_{\mu,600}(E,59°)}{S600(59°)} = 10^{1.81-1.74} = 1.17$ and $\frac{S^{exp}_{\mu,600}(E,57°)}{S600(57°)} = 10^{1.48-1.58} = 0.8$ correspondingly. Hence, the average value for these events is ≈1. This suggests that surface and underground detectors had registered virtually the same particles, namely – muons with energies above 1.87 GeV. This fact is hard to explain within the framework of modern understanding of possible nature of primary particles. Because even for iron nuclei, according to the simulation results presented on Fig. 2 and Fig. 4, the expected values are $\frac{S_{\mu,600}(E,59°,Fe)}{S600(59°)} = 10^{1.66-1.74} = 0.83$ and $\frac{S_{\mu,600}(E,57°,Fe)}{S600(57°)} = 10^{1.37-1.58} = 0.62$, which give the average muon fraction ≈0.82. This value is 1.25 times lower than the one obtained in experiment.

## CONCLUSION

From the presented data it is clear that the Arian shower and the event from 02/04/2024 – two most powerful EAS events registered at the Yakutsk EAS array to this day – are characterized by abnormally high muon content, which is hard interpret within the established notion of the nature of CR. The possibility that here we are dealing with some new exotic particles cannot be ruled out. Plans are to continue investigating this problem in the region of lower EAS energies.

## FUNDING

This work, except the section "*Event from 02/04/2024*", was conducted with the financial support of the Ministry of science and higher education of the Russian Federation (project № 0297-2021-011). The study described in section "*Event from 02/04/2024*" was supported by the grant of the Russian Science Foundation № 24-22-20048, https://rscf.ru/project/24-22-20048/ (grant executors A.A. Ivanov, K.G. Lebedev, S.V. Matarkin, I.S. Petrov).

## CONFLICT OF INTEREST

The authors of this work declare that they have no conflicts of interest.